\documentclass[10pt,pra,aps,onecolumn,superscriptaddress,floatfix,notitlepage]{revtex4-1}
\usepackage[latin1]{inputenc}
\usepackage{natbib}
\usepackage{amsmath}
\usepackage{amsfonts}
\usepackage{amssymb}
\usepackage{bm}
\usepackage{braket}
\usepackage[pdftex]{graphicx}
\usepackage{graphicx}
\usepackage{dblfloatfix}
\DeclareMathOperator{\sinc}{sinc}

\begin{document}
\title{Diffraction-Based Interaction-Free Measurements}
\author{Spencer Rogers}
\email{sroge22@ur.rochester.edu}
\affiliation{Department of Physics and Astronomy, University of Rochester, Rochester, NY 14627, USA}
\affiliation{Center for Coherence and Quantum Optics, University of Rochester, Rochester, NY 14627, USA}
\author{Yakir Aharonov}
\affiliation{School of Physics and Astronomy, Tel Aviv University, Tel Aviv 6997801, Israel}
\affiliation{Institute for Quantum Studies, Chapman University, Orange, CA 92866, USA}
\author{Cyril Elouard}
\affiliation{Department of Physics and Astronomy, University of Rochester, Rochester, NY 14627, USA}
\affiliation{Center for Coherence and Quantum Optics, University of Rochester, Rochester, NY 14627, USA}
\author{Andrew N. Jordan}
\affiliation{Department of Physics and Astronomy, University of Rochester, Rochester, NY 14627, USA}
\affiliation{Center for Coherence and Quantum Optics, University of Rochester, Rochester, NY 14627, USA}    
\affiliation{Institute for Quantum Studies, Chapman University, Orange, CA 92866, USA}
\date{\today}
\begin{abstract}
We introduce diffraction-based interaction-free measurements.  In contrast with previous work where a set of discrete paths is engaged, good quality interaction-free measurements can be realized with a continuous set of paths, as is typical of optical propagation.  If a bomb is present in a given spatial region - so sensitive that a single photon will set it off - its presence can still be detected without exploding it.  This is possible because, by not absorbing the photon, the bomb causes the single photon to diffract around it.  The resulting diffraction pattern can then be statistically distinguished from the bomb-free case.  We work out the case of single- versus double- slit in detail, where the double-slit arises because of a bomb excluding the middle region.
\end{abstract}
\maketitle

\section{Introduction}
The first example of a quantum interaction-free measurement (IFM) was given by Elitzur and Vaidman in 1993 \cite{elitzur1993quantum}. Their famous ``bomb tester'' is a Mach-Zehnder interferometer which may or may not have a malicious object (``bomb'') blocking one interferometer arm. In the absence of the blockage, one output detector is ``dark'' (never clicks) due to destructive interference of the two interferometer arms. If the blockage is present, interference between the two arms is lost, and the previously dark detector has some chance of clicking. Those photons that hit the dark detector inform the observer that the blockage is present, although such photons must have avoided the blockage in order to arrive at the detector in the first place. Thus, the blockage's prescence may be ascertained in an ``interaction-free'' manner. Since Elitzur and Vaidman's original proposal, strides have been made to improve the efficiency using the quantum Zeno Effect \cite{kwiat1995interaction}, achieve counterfactual communication (whereby two parties exchange information without exchanging particles) \cite{salih2013protocol,aharonov2019modification}, image in an interaction-free manner \cite{white1998interaction,zhang2019interaction}, compute the outcome of a quantum gate without running the gate (counterfactual computation) \cite{mitchison2001counterfactual,hosten2006counterfactual,kong2015experimental}, and gauge the extent to which these protocols may be considered truly ``interaction-free'' \cite{vaidman2003meaning,simon2000fundamental,dicke1986observing}. 

From the work of Feynman \cite{feynman2010quantum,feynman2006qed}, we know that probabilities in quantum mechanics may be seen as arising from a summation of probability amplitudes over histories with identical initial and final boundary conditions. In an interaction-free measurement, histories that meet some intermediate condition (i.e. those in which a particle touches the bomb en route to the objective) are excluded, altering the summation of amplitudes and thereby unlocking new possibilities (i.e. causing previously dark detectors to fire). Given the generality of Feynman's sum over histories, interaction-free measurements should be possible in a variety of contexts. In this paper, we consider interaction-free measurements based on diffraction. The essential effect investigated here is that a bomb, by excluding possible paths a particle can take without blowing it up, causes the quantum wave to diffract, resulting in a new interference pattern which can be distinguished from the no-bomb case.

In section \ref{section:bombInducedDoubleSlitSection}, we describe a simple experiment with a bomb and a slit, in which the bomb can be detected in the IFM manner. Using a 1-D model, we study two possible measurements and their efficiencies. Section \ref{section:bombInducedDoubleSlitSection} also contains a brief subsection on how the photon's momentum changes in the measurement. In section \ref{section:roleOfTime}, we analyze the role of the time-response of the bomb, and show that it must not be perfect if the bomb is to have some finite chance of exploding. We discuss possible implications for the interaction-free nature of the measurement.

\section{Bomb-Induced Double-Slit}\label{section:bombInducedDoubleSlitSection}
\subsection{Position Measurements}
\begin{figure}
\centering
  \includegraphics[scale=0.6]{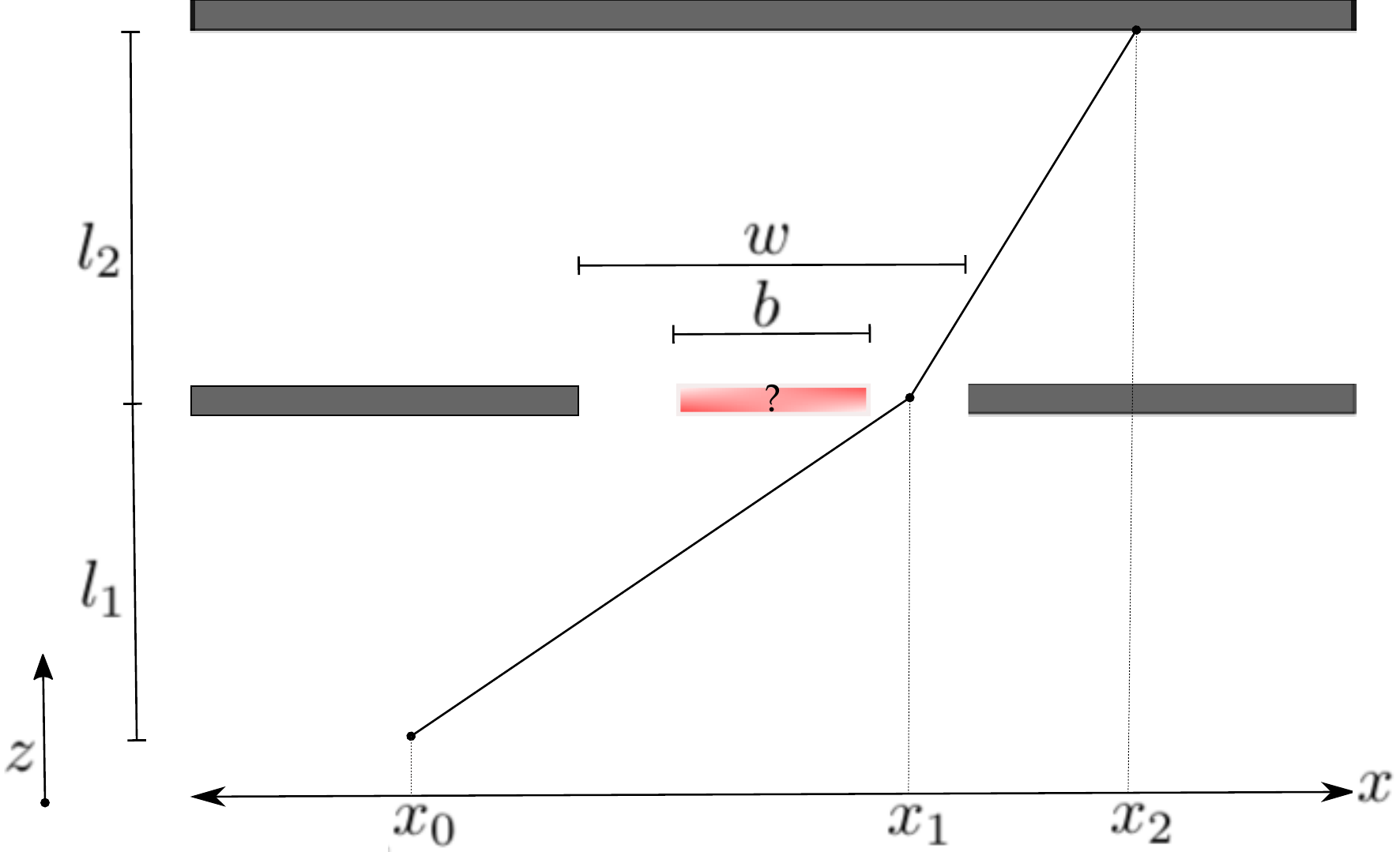}
  \caption{The probability that a photon reaches a point $x_2$ on the back screen is computed by a sum of probability amplitudes for the various paths leading to $x_2$. If the bomb (red) is present, paths through it are excluded from the summation, affecting the result and potentially allowing the photon to land on some point it otherwise could not. A similar diagram is used in Ref. \cite{feynman2010quantum}, page 48.} 
  \label{figure:schematic}
\end{figure}

The primary system we use to demonstrate diffraction-based IFM is a single photon moving toward a detection screen and an opaque screen with a slit, allowing the photon to pass. The slit may or may not contain a thin bar-shaped detector (``bomb'') in the middle of it (see Fig. \ref{figure:schematic}). This makes the apparatus act like a double-slit if the bomb is present, and a single slit if the bomb is not. We tune our apparatus so that the single-slit interference pattern that results if the bomb is not present contains zeros (points the photon cannot reach). These are analogous to the dark port in the Elitzur-Vaidman protocol. Putting the bomb in place gives the photon a chance to land on these zeros (while still avoiding the bomb). A detection at one of these zeros would be a sign that the bomb is definitely present and constitute an interaction-free measurement.

From a Bayesian perspective, photons that do not land on these zeros are also informative. Given some prior that the bomb is present $P(\mathrm{Bomb})$, finding the photon at a position $x$ causes our posterior belief in the bomb's presence to be (applying Bayes' rule):
\\
\begin{equation}
\label{eq:BayesRule}
    P(\mathrm{Bomb}|x)=\frac{P(x|\mathrm{Bomb})P(\mathrm{Bomb})}{P(x|\mathrm{Bomb})P(\mathrm{Bomb})+P(x|\mathrm{No} \: \mathrm{Bomb})P(\mathrm{No}\: \mathrm{Bomb})}.
\end{equation}
\\
$P(x|\mathrm{Bomb})$ and $P(x|\mathrm{No}\:\mathrm{Bomb})$ are the probability density functions for the photon to be found in $x$ conditional on the bomb being present and not present, respectively, and should behave as double- and single- slit intensities at $x$, respectively. The priors $P(\mathrm{Bomb})$ and $P(\mathrm{No}\:\mathrm{Bomb})$ sum to 1; for simplicity, we assume the person using the bomb-tester knows (or, less strongly, believes) that either the bomb is present in the middle of the slit with known dimensions, or is not present. We could certainly analyze situations where the observer using the bomb-tester is less certain of the bomb's potential position and dimensions, but we refrain from doing so here so as not to detract from the primary aims of the manuscript. That the photon provides information regardless of whether it lands exactly on a zero is important, since the probability of landing on this region of zero area is zero. From a practical standpoint, we care that measurement results are capable of raising our posterior satisfactorily close to 1, so as to have a confident diagnosis of the bomb being present.

We model the photon's transverse degree of freedom as a 1-D quantum system subject to the paraxial free-space Hamiltonian $\hat{\mathcal{H}}=\frac{\hbar\hat{k}_x^2}{2k_0}c$, where $k_0$ is the wavenumber in the propagation direction $\hat{z}$, $x$ is the transverse coordinate, and $\hat{k}_x=-i{\frac{\hat{\partial}}{\partial x}}$ is the scaled momentum operator in the transverse direction. Associated to this Hamiltonian is the evolution operator $\hat{U}_l=e^{-i\hat{k}_x^2l/2k_0}$, where $l$ is the distance traveled by the photon in the $z$-direction \cite{dressel2013strengthening}. The matrix element $\bra{x'}\hat{U}_l\ket{x}=\sqrt{\frac{k_0}{2\pi il}}e^{\frac{ik_0(x'-x)^2}{2l}}$ is the probability amplitude for the photon to travel from $x$ to $x'$ (as it travels $l$ in the $z$-direction). We may think of the photon's journey as consisting of two parts. In the first part, the photon travels a distance $l_1$ in the $z$-direction and arrives at the plane of the slit. If the photon reaches an opening in the screen and avoids the bomb, the photon's journey has a second part of interest to us, in which the photon travels a distance $l_2$ in the $z$-direction to a detection screen. The slit, bomb, and detection screen are all parallel to the $x$-axis (see Fig. \ref{figure:schematic}).

A \textit{history} is a sequence of coordinates $(x_0,x_1,x_2)$ that specifies where the photon starts, passes the slit plane, and arrives at the detection screen. Each history has associated with it some probability amplitude $K(x_0,x_1,x_2)=\bra{x_2}\hat{U}_{l_2}\ket{x_1}\bra{x_1}\hat{U}_{l_1}\ket{x_0}\braket{x_0|\Psi}$, where $\ket{\Psi}$ is the photon's transverse quantum state. The total probability amplitude for arriving at $x_2$ is obtained by integrating the probability amplitudes for all possible histories ending in $x_2$. If the bomb is not present, the total probability amplitude is
\begin{equation}
\label{eq:PsiNoBomb}
\braket{x_2|\Psi}_{\mathrm{No}\:\mathrm{Bomb}}=\int^{w/2}_{-w/2} dx_1 \int^{\infty}_{-\infty} dx_0 \bra{x_2}\hat{U}_{l_2}\ket{x_1}\bra{x_1}\hat{U}_{l_1}\ket{x_0}\braket{x_0|\Psi},
\end{equation}
and the probability density function for reaching $x_2$ is $P(x_2|\mathrm{No}\: \mathrm{Bomb})=|\braket{x_2|\Psi}_{\mathrm{No}\:\mathrm{Bomb}}|^2$. $w$ is the width of the slit. If the bomb is present, the result is
\begin{equation}
\label{eq:PsiBomb}
\begin{split}
\braket{x_2|\Psi}_{\mathrm{Bomb}}&=\braket{x_2|\Psi}_{\mathrm{No}\:\mathrm{Bomb}}-\int^{b/2}_{-b/2} dx_1 \int^{\infty}_{-\infty} dx_0 \bra{x_2}\hat{U}_{l_2}\ket{x_1}\bra{x_1}\hat{U}_{l_1}\ket{x_0}\braket{x_0|\Psi},\\
&=\braket{x_2|\Psi}_{\mathrm{No}\:\mathrm{Bomb}}-\braket{x_2|\Psi}_{\mathrm{Excluded}},
\end{split}
\end{equation}
where the second term refers to histories that involve the photon passing through the bomb (these are excluded since they would cause the bomb to detect the photon and explode), and $b$ is the length of the bomb. The probability density function for reaching $x_2$ in this case is $P(x_2|\mathrm{Bomb})=|\braket{x_2|\Psi}_{\mathrm{Bomb}}|^2$. We note that, as written in Eq. \ref{eq:PsiNoBomb} and \ref{eq:PsiBomb}, $\braket{x_2|\Psi}_{\mathrm{No}\:\mathrm{Bomb}}$ and $\braket{x_2|\Psi}_{\mathrm{Bomb}}$ are un-normalized wavefunctions. $\int^{\infty}_{-\infty}dx_2\:|\braket{x_2|\Psi}_{\mathrm{No}\:\mathrm{Bomb}}|^2$ is the probability that the photon reaches the back screen if the bomb is not present, and $\int^{\infty}_{-\infty}dx_2\:|\braket{x_2|\Psi}_{\mathrm{Bomb}}|^2$ is the probability that the photon reaches the back screen if the bomb is present.

The probabilities depend on the initial position-space wavefunction $\braket{x_0|\Psi}$. In general, the integrals in Eq. \ref{eq:PsiNoBomb} and \ref{eq:PsiBomb} will be difficult to evaluate. The simplest case is when $l_1$ is very long, so that the wavefunction has time to broaden and become effectively constant over the slit ($\int^{\infty}_{-\infty}dx_0 \bra{x_1}\hat{U}_{l_1}\ket{x_0}\braket{x_0|\Psi}\simeq Z$ for all $x_1 \in (-w/2, w/2)$ and some complex constant $Z$). This is effectively the plane wave limit. In this limit, Eq. \ref{eq:PsiNoBomb} reduces to

\begin{equation}
\begin{split}
    \braket{x_2|\Psi}_{\mathrm{No}\:\mathrm{Bomb}}&=Z\int^{w/2}_{-w/2} dx_1 \:\sqrt{\frac{k_0}{2\pi il_2}}e^{\frac{ik_0(x_1-x_2)^2}{2l_2}},\\
    &=\tilde{Z}\int^{\sqrt{\frac{k_0}{\pi l_2}}(\frac{w}{2}-x_2)}_{\sqrt{\frac{k_0}{\pi l_2}}(-\frac{w}{2}-x_2)}dt\:e^{i\pi t^2/2},\\
    &=\tilde{Z}\:\bigg\{\mathcal{C}\Big[\sqrt{\frac{k_0}{\pi l_2}}(\frac{w}{2}-x_2)\Big]-\mathcal{C}\Big[\sqrt{\frac{k_0}{\pi l_2}}(-\frac{w}{2}-x_2)\Big]\\
    &\:\:\:\:\:+i\mathcal{S}\Big[\sqrt{\frac{k_0}{\pi l_2}}(\frac{w}{2}-x_2)\Big]-i\mathcal{S}\Big[\sqrt{\frac{k_0}{\pi l_2}}(-\frac{w}{2}-x_2)\Big]\bigg\},
    \label{eq:FresnelResult}
\end{split}
\end{equation}
where empirically irrelevant normalization factors have been absorbed into $\tilde{Z}$. $\mathcal{C}$ and $\mathcal{S}$ are the Fresnel integrals: $\mathcal{C}(x)=\int^x_0 dt \cos{\frac{\pi t^2}{2}}$ and $\mathcal{S}(x)=\int^x_0 dt \sin{\frac{\pi t^2}{2}}$ \cite{goodman2005introduction}. The second term in Eq. \ref{eq:PsiBomb} is of the same form, and can thus be evaluated the same way. Similar calculations occur in Ref. \cite{sawant2014nonclassical} and Ref. \cite{beau2012feynman}.

\begin{figure}
\centering
  \includegraphics[scale=0.42]{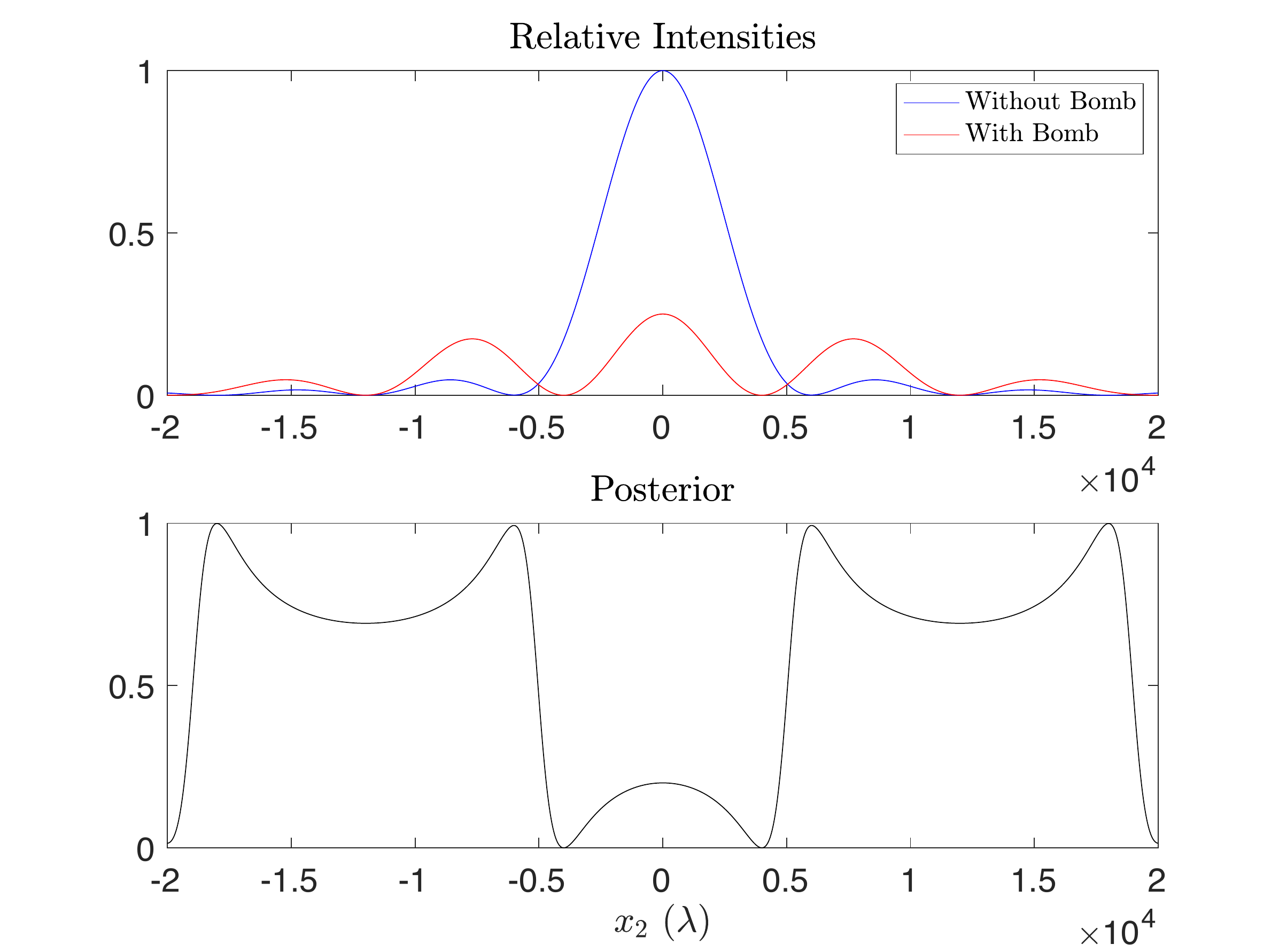}
  \caption{Intensities at the back screen in units such that $I(0)=1$ without the bomb (top). Posterior belief in the bomb's presence given a single photon detection at $x_2$ and a prior of 0.5 (bottom). The lengths chosen were, in units of $\lambda=\frac{2\pi}{k_0}$, $w=1000$, $b=500$, and $l_2=6 \times 10^6$. The plane wave limit described in Eq. \ref{eq:FresnelResult} was used (and since $\frac{b}{w}=0.5$, the bomb should explode 50\% of the time the photon reaches the slit). For IFM purposes, the most important values of $x_2$ are those for which the single-slit pattern (top, blue) goes to zero; at these points, the posterior goes to one.}
  \label{figure:data}
\end{figure}

Sample results for our bomb-tester are shown in Fig. \ref{figure:data}. The results demonstrate clearly that there exist points $x_2$ with both zero probability amplitude when the bomb is not present and nonzero probability amplitude when the bomb is present, provided one tunes the lengths in the problem appropriately. This suffices to show that there is an analogue of the dark port of the original Elitzur-Vaidman bomb-tester present in our protocol.

The standard efficiency measure for an IFM protocol is $\eta=\frac{P(\mathrm{Dark}|\mathrm{Bomb})}{P(\mathrm{Dark}|\mathrm{Bomb})+P(\mathrm{Explosion}|\mathrm{Bomb})}$ \cite{kwiat1995interaction}. However, since the dark ports of our protocol are points in a continuum, this measure gives zero. Instead we define a measure based on how likely a single photon detection is to cause our posterior (Eq. \ref{eq:BayesRule}) to exceed some threshold $T$:

\begin{equation}
    \tilde{\eta}(T,P(\mathrm{Bomb}))=\frac{\int^{\infty}_{-\infty}dx_2\: P(x_2|\mathrm{Bomb})\Theta(P(\mathrm{Bomb}|x_2)-T)}
    {
    \int^{\infty}_{-\infty}dx_2\: P(x_2|\mathrm{Bomb})\Theta(P(\mathrm{Bomb}|x_2)-T)+P(\mathrm{Explosion}|\mathrm{Bomb})
    },
\end{equation}
where $\Theta$ is the Heaviside function. This is, notably, dependent on both prior and threshold. For the parameters in Fig. \ref{figure:data}, a prior $P(\mathrm{Bomb})=0.5$, and a threshold $T=0.99$, we calculated $\tilde{\eta} \approx 2.1\%$. This corresponds to about 47 bombs exploding per one detected in the IFM manner (where by `detected' we mean the observer believes the bomb exists with at least 99\% certainty).

One may think of this section as having described a sort of classification algorithm: if a single-photon position measurement causes our posterior belief in the bomb's presence to be greater than a certain threshold, classify the slit as containing the bomb. We likewise classify a slit as not containing a bomb if a single-photon test causes our posterior belief in the bomb's absence to be greater than the same threshold. If a single photon hits the back screen but does not cause our posterior to be in either of these extremes, the test is inconclusive. For $T\neq 1$, our classifier \textit{will} make some mistakes, classifying empty slits as bomb-containing and bomb-containing slits as empty. We can test our classification algorithm using rejection sampling to see how accurate our classifier is.

At this point, we are not so much interested in how often the bomb explodes relative to how often we obtain useful results. We are more interested in how accurately our classifier tells a regular double-slit and single-slit apart. This is still a two-hypothesis test, where the dimensions of the bodies are known. Drawing samples from the PDFs of Fig.  \ref{figure:data}, with x-axes expanded out to $\pm 10^6\lambda$ and bins of width 10$\lambda$ (a typical region where the posterior exceeds 99\% is then about forty bins wide), we subjected double-slits and single-slits to single-photon tests. The classifier, using a threshold of 99\%, successfully classified 25,625 (2.14\%) of the 1,197,851 double-slits. 266 double-slits were incorrectly classified as single-slits. The remaining double-slits were each probed by a photon but the results were inconclusive. 98.97\% of the classifier's claims were accurate in this case. Of 600,439 single-slits, 16,494 (2.75\%) were correctly classified, 42 were incorrectly classified as double-slits, and the remaining slits tested inconclusively. 99.75\% of the classifier's claims were accurate in this case. This leads us to estimate that if equal numbers of double- and single-slits were tested, the classifier would be correct about 99.7\% of the times that it claims to see a double-slit. By the same reasoning, it would be correct about 99.2\% of the times that it claims to see a single-slit.

Lastly, we would like to clarify that while we can correctly classify single-slits as not containing bombs with this protocol, this is not done in the IFM manner (the proposed bomb region is probed by photons which cannot be said to have avoided that region).

\subsection{Optimal Measurements}
The full quantum formalism describes the results of measurements other than position measurements. Using these, the bomb-induced double-slit may, in theory, be made as efficient as the original Elitzur-Vaidman bomb-tester. To see this explicitly, consider the wavefunction of the photon at the slit, $\braket{x_1|\Psi}$. As previously stated, it is effectively constant over the slit in the plane wave limit (unless the bomb is present). If the bomb is present, and does not explode, the wavefunction collapses to zero where the bomb is, and is constant over the rest of the slit.

\begin{align}
\braket{x_1|\Psi}_{\mathrm{No}\:\mathrm{Bomb}}&=
 \begin{cases}
 \sqrt{\frac{1}{w}} & |x|\leq w/2,\\
 0 & |x|>w/2,
 \end{cases}\label{eq:conditional1}\\
\braket{x_1|\Psi}_{\mathrm{Bomb}}&=     \begin{cases}
      0 & |x| < b/2, |x|>w/2, \\
      \sqrt{\frac{1}{w-b}} & b/2\leq |x|\leq w/2. \\
   \end{cases}
\label{eq:conditional2}
\end{align}
For clarification, Eq. \ref{eq:conditional1} and \ref{eq:conditional2} use a different convention from that of section 2.1. Eq. \ref{eq:conditional1} and \ref{eq:conditional2} describe \textbf{normalized} wavefunctions; they assume it is known that the photon is in the slit and fails to set off any bomb which might be there. By contrast, $\braket{x_2|\Psi}_{\mathrm{No}\:\mathrm{Bomb}}$ and $\braket{x_2|\Psi}_{\mathrm{Bomb}}$ of Eq. \ref{eq:PsiNoBomb} and \ref{eq:PsiBomb} are not normalized wavefunctions; they are reduced because the photon has a probability of hitting the opaque part of the screen and a probability of setting off the bomb ($\braket{x_2|\Psi}_{\mathrm{No}\:\mathrm{Bomb}}$ has a larger norm than $\braket{x_2|\Psi}_{\mathrm{Bomb}}$ in Eq. \ref{eq:PsiNoBomb} and \ref{eq:PsiBomb}).

\begin{figure}[b]
\centering
  \includegraphics[scale=0.7]{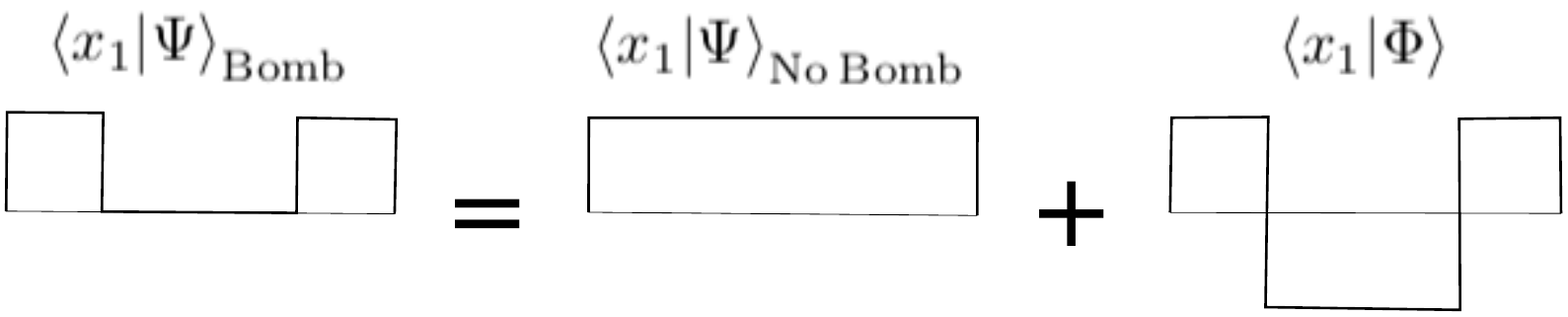}
  \caption{The wavefunction at the slit, if the bomb is present, is a sum of two orthogonal terms. The effect of the bomb (when it does not explode) is to cause $\ket{\Psi}_{\mathrm{No}\:\mathrm{Bomb}}$ to collapse to $\ket{\Psi}_{\mathrm{Bomb}}$. This gives the state a nonzero projection onto $\ket{\Phi}$, which it did not have previously.}
  \label{figure:superposition}
\end{figure}

There is a third wavefunction of interest:
\begin{align}
\braket{x_1|\Phi}&=
 \begin{cases}
 -\sqrt{\frac{w-b}{wb}} & |x| < b/2,\\
 \sqrt{\frac{b}{w(w-b)}} & b/2\leq |x|\leq w/2,\\
 0 & |x|>w/2.
 \end{cases}\label{eq:conditional3}
\end{align}
$\ket{\Phi}$ is like the dark port in the Elitzur-Vaidman Bomb-tester. It is orthogonal to the wavefunction that results if the bomb is not present: $\int^{\infty}_{-\infty}dx_1 \braket{\Phi|x_1}\braket{x_1|\Psi}_{\mathrm{No}\:\mathrm{Bomb}}=0$, but not to the wavefunction that results if the bomb is present: $\int^{\infty}_{-\infty}dx_1 \braket{\Phi|x_1}\braket{x_1|\Psi}_{\mathrm{Bomb}}=\sqrt{\frac{b}{w}}$. The wavefunction $\braket{x_1|\Psi}_{\mathrm{Bomb}}$ is simply the superposition of the other two wavefunctions (see Fig. \ref{figure:superposition}):

\begin{equation}
    \braket{x_1|\Psi}_{\mathrm{Bomb}}= \sqrt{\frac{w-b}{w}}\braket{x_1|\Psi}_{\mathrm{No}\:\mathrm{Bomb}} + \sqrt{\frac{b}{w}}\braket{x_1|\Phi}.
\end{equation}
This holds for later times as well: $\braket{x_2|\Psi}_{\mathrm{Bomb}}= \sqrt{\frac{w-b}{w}}\braket{x_2|\Psi}_{\mathrm{No}\:\mathrm{Bomb}} + \sqrt{\frac{b}{w}}\braket{x_2|\Phi}$, since unitary evolution preserves inner products.

In the optimal IFM protocol, one measures whether the photon is in the state $\ket{\Psi}_{\mathrm{No}\:\mathrm{Bomb}}$ or $\ket{\Phi}$ (Ref. \cite{broadbent2009discriminating} discusses holographic techniques which might be applied to perform this). If the photon is found in the state $\ket{\Phi}$, one concludes that the bomb is present, since the result would be impossible otherwise. Conditioned on the photon arriving at the slit opening and the bomb not exploding, the result $\ket{\Psi}_{\mathrm{No}\:\mathrm{Bomb}}$ occurs with probability $\frac{w-b}{w}$ and the result $\ket{\Phi}$ occurs with probability $\frac{b}{w}$. The IFM efficiency is

\begin{equation}
\eta=\frac{P(\ket{\Phi}|\mathrm{Bomb})}{P(\ket{\Phi}|\mathrm{Bomb})+P(\mathrm{Explosion}|\mathrm{Bomb})}=\frac{(1-\frac{b}{w})\frac{b}{w}}{(1-\frac{b}{w})\frac{b}{w}+\frac{b}{w}}=\frac{1-\frac{b}{w}}{2-\frac{b}{w}},
\end{equation}
for $0 < \frac{b}{w} < 1$. As a function of $\frac{b}{w}$, $\eta$ has no maximum, but approaches the lowest upper bound $\frac{1}{2}$ as $\frac{b}{w}\rightarrow 0$. This corresponds to a situation in which $P(\ket{\Phi})\approx P(\mathrm{Explosion}) = \frac{b}{w} \ll 1$ (conditioned on the photon making it to the slit opening in the first place); $\ket{\Phi}$ and explosions are both unlikely, but equally so. Naturally, $\eta \rightarrow 0$ as $\frac{b}{w}\rightarrow 1$, since $P(\mathrm{Explosion})\rightarrow 1$ and $P(\ket{\Phi})\rightarrow 0$. In the original Elitzur-Vaidman bomb-tester, the effiency was $\eta=\frac{1-R}{2-R}$, where $R$ is the reflectivity of the beam splitters used \cite{kwiat1995interaction}, so that the efficiencies of the two protocols are equal if $\frac{b}{w}=R$.

\subsection{The Momentum Picture}

\begin{figure}[b]
\centering
  \includegraphics[scale=0.5]{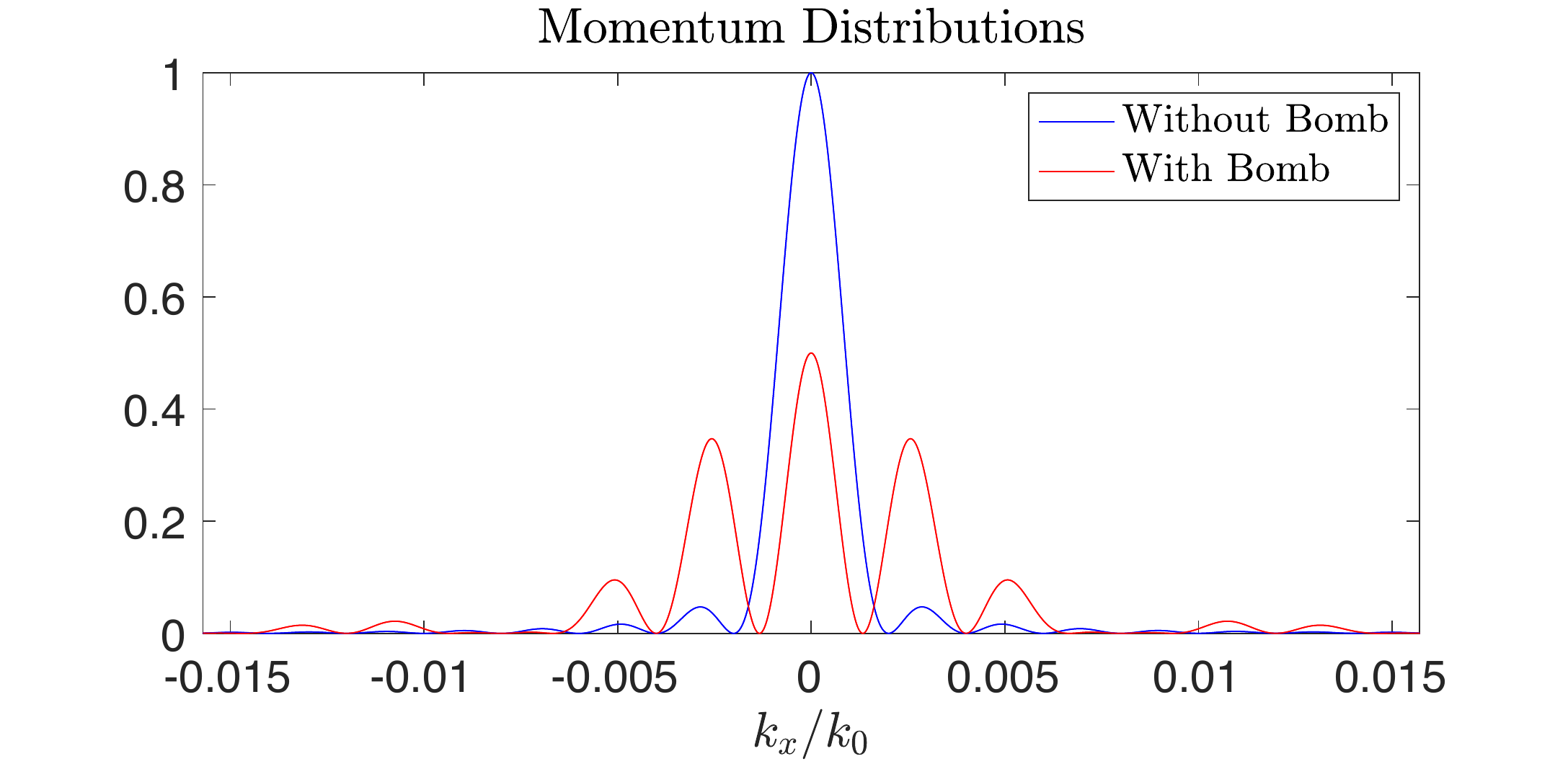}
  \caption{The momentum distribution which results if the bomb is present (and does not explode) is distinct from the momentum distribution that results if the bomb is not present. The bomb ``kicks'' the photon in this sense. Due to symmetry, the average momentum is zero regardless. However, the with-bomb distribution tends to yield a greater magnitude $|k_x|$. The PDFs are normalized so that the no-bomb distribution is $1$ at $k_x=0$. The transverse momentum is given in units of the $z$-momentum $k_0$. Parameters are those used in Fig. \ref{figure:data}.}
  \label{figure:momentum}
\end{figure}

When the bomb does not explode, and the position-space wavefunction updates from that of Eq. \ref{eq:conditional1} to that of Eq. \ref{eq:conditional2}, the momentum-space wavefunction also changes. The Fourier transforms of these piecewise constant wavefunctions are easily computed:

\begin{equation}
    \braket{k_x|\Psi}_{\mathrm{No}\:\mathrm{Bomb}}=\sqrt{\frac{w}{2\pi}}\sinc{\big(\frac{k_xw}{2}\big)}
    \label{eq:fourTrans1}
\end{equation}
and
\begin{equation}
    \braket{k_x|\Psi}_{\mathrm{Bomb}}=\sqrt{\frac{1}{2\pi(w-b)}}\Bigg[w\sinc{\big(\frac{k_xw}{2}}\big)-b\sinc{\big(\frac{k_xb}{2}}\big)\Bigg].
    \label{eq:fourTrans2}
\end{equation}
The momentum distributions corresponding to these two wavefunctions, $|\braket{k_x|\Psi}_{\mathrm{No}\:\mathrm{Bomb}}|^2$ and $|\braket{k_x|\Psi}_{\mathrm{Bomb}}|^2$, are shown in Figure \ref{figure:momentum}. Like the dark ports in the Elitzur-Vaidman protocol, there are transverse momenta that become accessible when the bomb is present (there are zeros of the no-bomb distribution which are not zeros of the bomb distribution). In this sense, the bomb ``kicks'' the photon, even when the photon avoids it. We note that this momentum change is a feature of any measurement protocol which checks if the photon of Eq. \ref{eq:conditional1} is in the region $(-b/2, b/2)$. Without a detailed Hamiltonian description of the measurement procedure used, the mechanism by which the momentum changes is unspecified.

Momentum-conserving scattering traditionally occurs via repulsive interaction potentials that depend on the distances between the involved objects. The effect of such potentials on the photon is not investigated here, but we note that the bomb's momentum should be sufficiently uncertain that interference of the two open slits is not lost in the presence of such a potential. This is important since such a potential would shift the bomb momentum to the right if the photon passes on the left, and shift the bomb momentum to the left if the photon passes on the right; a high degree of entanglement would result if the bomb momentum were not sufficiently uncertain, destroying the double-slit interference pattern. The single-slit patterns we would obtain if the bomb's momentum were too precise would not necessarily make IFM impossible, but our double-slit calculation would be inapplicable.

\section{The Role of Time}
\label{section:roleOfTime}
Throughout this paper, we have treated the bomb as though it interacts with the photon at a specific instant; the instant the photon reaches the slit plane, the bomb checks if the photon is in its space, and explodes accordingly. This treatment suffices to demonstrate the essential concepts of diffraction-based IFM, but ignores certain interesting phenomena.

Had we modeled the photon as a 2-D or 3-D quantum system, we would have had to account for the fact that its wavefunction technically has non-vanishing tails. In particular, the photon wavefunction is nonzero where the bomb is (at, essentially, all times). The situation is one in which the photon and bomb are constantly coupled, determining whether or not the bomb should explode based on how much the photon wavefunction overlaps with the bomb, and updating the photon wavefunction accordingly (zeroing it out temporarily in the bomb region if the bomb does not explode).

For concreteness, suppose we model the bomb as a measuring device which checks if the photon is in its space at equal intervals $\Delta t$. If the bomb does not find the photon, the photon wavefunction goes to zero in the bomb region. Before the next measurement, the photon has some time to develop an amplitude in the bomb region. If $\Delta t$ is very short, however, this amplitude will be small. In fact, the probability that the photon is ever found in the bomb region goes to zero as $\Delta t \rightarrow 0$. This is a manifestation of the quantum Zeno effect, whereby a quantum system is prevented from leaving an eigenspace of an observable by the constant measurement of that observable \cite{misra1977zeno}. A number of sources have noted this effect before; a detector which measures too often acts like an infinite potential barrier \cite{mackrory2010reflection,halliwell2012pitfalls}. Our treatment in section 2 ignores slit and bomb thickness. If these have a thickness $\delta$, then $\delta/c$ becomes a relevant timescale in our problem, since this is the natural time it takes the photon to pass. To absorb the photon with the ``common-sense'' probability, the bomb should measure at intervals $\Delta t < \delta/c$ but \textbf{not} $\Delta t \ll \delta/c$. This way, complications due to the Zeno effect are avoided.

We have seen that, in order for the bomb to explode at all, its measurement rate must be finite. This leaves open the possibility for the photon to pass through the bomb \textit{between} measurements (e.g. on timescales much shorter than $\delta/c$) without setting it off. In the bomb-induced double-slit, if $\delta \ll b$, an overwhelming majority of paths through the bomb will be of length $\delta$ or greater (too long for speed-$c$ photons to pass undetected), so that the contribution from these ``undetected photon'' histories will have little to no effect on the interference pattern. However, the claim that the photon definitely avoided the bomb region is affected; the bomb not exploding is consistent with the photon passing through it on some short time scale. Lastly, we note that this problem is not unique to diffraction-based IFM, since the bombs in other IFM protocols must also be deficient at some level in order to have some chance of exploding (this point was realized differently by the authors of \cite{simon2000fundamental}, who noted that the bomb should not be sensitive to momentum kicks on the order of its own quantum fluctuations $\Delta p$).

\section{Conclusion}
We have shown how interaction-free measurements can be applied to systems with a continuum of paths. An effect present in such systems is that, by measuring the particle's position, the bomb naturally alters the particle's momentum state. In the particular case of the bomb-induced double-slit, we showed that  the IFM efficiency can match that of the original Elitzur-Vaidman bomb-tester, provided one chooses the appropriate measurement basis. While we have focused in this paper on a bomb with a predetermined and known shape, we stress the overall conclusions may be generalized to a bomb of arbitrary shape.  The presence of such a bomb may be detected, provided it does not explode, if a prepared photon experiences an anomalously large momentum kick, such that further propagation causes the photon to register at an unlikely position on the screen, relative to the initial waveform of the prepared photon. We have also considered the role of the finite time-response of the bomb, which affects the claim that the probe particle completely avoided the bomb region. However, finite time-response of the bomb does not prevent us from detecting the bomb without exploding it, and is relevant in other IFM protocols as well.  Thus, diffraction-based interaction-free measurements are as interaction-free as previous interaction-free measurements.

\section*{Acknowledgements}
We thank John Howell, Paul Kwiat, and Jeff Tollaksen for helpful discussions.  This work was supported by the NSF grant DMR-1809343.  Y.A. acknowledges support from the Israel Science Foundation (Grant 1311/14), Israeli Centers of Research Excellence (ICORE) Center ``Circle of Light,'' and the German-Israeli Project Cooperation (Deutsch-Israelische Projektkooperation, DIP).

\bibliography{ref.bib}

\begin{thebibliography}{22}%
\makeatletter
\providecommand \@ifxundefined [1]{%
 \@ifx{#1\undefined}
}%
\providecommand \@ifnum [1]{%
 \ifnum #1\expandafter \@firstoftwo
 \else \expandafter \@secondoftwo
 \fi
}%
\providecommand \@ifx [1]{%
 \ifx #1\expandafter \@firstoftwo
 \else \expandafter \@secondoftwo
 \fi
}%
\providecommand \natexlab [1]{#1}%
\providecommand \enquote  [1]{``#1''}%
\providecommand \bibnamefont  [1]{#1}%
\providecommand \bibfnamefont [1]{#1}%
\providecommand \citenamefont [1]{#1}%
\providecommand \href@noop [0]{\@secondoftwo}%
\providecommand \href [0]{\begingroup \@sanitize@url \@href}%
\providecommand \@href[1]{\@@startlink{#1}\@@href}%
\providecommand \@@href[1]{\endgroup#1\@@endlink}%
\providecommand \@sanitize@url [0]{\catcode `\\12\catcode `\$12\catcode
  `\&12\catcode `\#12\catcode `\^12\catcode `\_12\catcode `\%12\relax}%
\providecommand \@@startlink[1]{}%
\providecommand \@@endlink[0]{}%
\providecommand \url  [0]{\begingroup\@sanitize@url \@url }%
\providecommand \@url [1]{\endgroup\@href {#1}{\urlprefix }}%
\providecommand \urlprefix  [0]{URL }%
\providecommand \Eprint [0]{\href }%
\providecommand \doibase [0]{http://dx.doi.org/}%
\providecommand \selectlanguage [0]{\@gobble}%
\providecommand \bibinfo  [0]{\@secondoftwo}%
\providecommand \bibfield  [0]{\@secondoftwo}%
\providecommand \translation [1]{[#1]}%
\providecommand \BibitemOpen [0]{}%
\providecommand \bibitemStop [0]{}%
\providecommand \bibitemNoStop [0]{.\EOS\space}%
\providecommand \EOS [0]{\spacefactor3000\relax}%
\providecommand \BibitemShut  [1]{\csname bibitem#1\endcsname}%
\let\auto@bib@innerbib\@empty
\bibitem [{\citenamefont {Elitzur}\ and\ \citenamefont
  {Vaidman}(1993)}]{elitzur1993quantum}%
  \BibitemOpen
  \bibfield  {author} {\bibinfo {author} {\bibfnamefont {A.~C.}\ \bibnamefont
  {Elitzur}}\ and\ \bibinfo {author} {\bibfnamefont {L.}~\bibnamefont
  {Vaidman}},\ }\href@noop {} {\bibfield  {journal} {\bibinfo  {journal}
  {Foundations of Physics}\ }\textbf {\bibinfo {volume} {23}},\ \bibinfo
  {pages} {987} (\bibinfo {year} {1993})}\BibitemShut {NoStop}%
\bibitem [{\citenamefont {Kwiat}\ \emph {et~al.}(1995)\citenamefont {Kwiat},
  \citenamefont {Weinfurter}, \citenamefont {Herzog}, \citenamefont
  {Zeilinger},\ and\ \citenamefont {Kasevich}}]{kwiat1995interaction}%
  \BibitemOpen
  \bibfield  {author} {\bibinfo {author} {\bibfnamefont {P.}~\bibnamefont
  {Kwiat}}, \bibinfo {author} {\bibfnamefont {H.}~\bibnamefont {Weinfurter}},
  \bibinfo {author} {\bibfnamefont {T.}~\bibnamefont {Herzog}}, \bibinfo
  {author} {\bibfnamefont {A.}~\bibnamefont {Zeilinger}}, \ and\ \bibinfo
  {author} {\bibfnamefont {M.~A.}\ \bibnamefont {Kasevich}},\ }\href@noop {}
  {\bibfield  {journal} {\bibinfo  {journal} {Physical Review Letters}\
  }\textbf {\bibinfo {volume} {74}},\ \bibinfo {pages} {4763} (\bibinfo {year}
  {1995})}\BibitemShut {NoStop}%
\bibitem [{\citenamefont {Salih}\ \emph {et~al.}(2013)\citenamefont {Salih},
  \citenamefont {Li}, \citenamefont {Al-Amri},\ and\ \citenamefont
  {Zubairy}}]{salih2013protocol}%
  \BibitemOpen
  \bibfield  {author} {\bibinfo {author} {\bibfnamefont {H.}~\bibnamefont
  {Salih}}, \bibinfo {author} {\bibfnamefont {Z.-H.}\ \bibnamefont {Li}},
  \bibinfo {author} {\bibfnamefont {M.}~\bibnamefont {Al-Amri}}, \ and\
  \bibinfo {author} {\bibfnamefont {M.~S.}\ \bibnamefont {Zubairy}},\
  }\href@noop {} {\bibfield  {journal} {\bibinfo  {journal} {Physical review
  letters}\ }\textbf {\bibinfo {volume} {110}},\ \bibinfo {pages} {170502}
  (\bibinfo {year} {2013})}\BibitemShut {NoStop}%
\bibitem [{\citenamefont {Aharonov}\ and\ \citenamefont
  {Vaidman}(2019)}]{aharonov2019modification}%
  \BibitemOpen
  \bibfield  {author} {\bibinfo {author} {\bibfnamefont {Y.}~\bibnamefont
  {Aharonov}}\ and\ \bibinfo {author} {\bibfnamefont {L.}~\bibnamefont
  {Vaidman}},\ }\href@noop {} {\bibfield  {journal} {\bibinfo  {journal}
  {Physical Review A}\ }\textbf {\bibinfo {volume} {99}},\ \bibinfo {pages}
  {010103} (\bibinfo {year} {2019})}\BibitemShut {NoStop}%
\bibitem [{\citenamefont {White}\ \emph {et~al.}(1998)\citenamefont {White},
  \citenamefont {Mitchell}, \citenamefont {Nairz},\ and\ \citenamefont
  {Kwiat}}]{white1998interaction}%
  \BibitemOpen
  \bibfield  {author} {\bibinfo {author} {\bibfnamefont {A.~G.}\ \bibnamefont
  {White}}, \bibinfo {author} {\bibfnamefont {J.~R.}\ \bibnamefont {Mitchell}},
  \bibinfo {author} {\bibfnamefont {O.}~\bibnamefont {Nairz}}, \ and\ \bibinfo
  {author} {\bibfnamefont {P.~G.}\ \bibnamefont {Kwiat}},\ }\href@noop {}
  {\bibfield  {journal} {\bibinfo  {journal} {Physical Review A}\ }\textbf
  {\bibinfo {volume} {58}},\ \bibinfo {pages} {605} (\bibinfo {year}
  {1998})}\BibitemShut {NoStop}%
\bibitem [{\citenamefont {Zhang}\ \emph {et~al.}(2019)\citenamefont {Zhang},
  \citenamefont {Sit}, \citenamefont {Bouchard}, \citenamefont {Larocque},
  \citenamefont {Grenapin}, \citenamefont {Cohen}, \citenamefont {Elitzur},
  \citenamefont {Harden}, \citenamefont {Boyd},\ and\ \citenamefont
  {Karimi}}]{zhang2019interaction}%
  \BibitemOpen
  \bibfield  {author} {\bibinfo {author} {\bibfnamefont {Y.}~\bibnamefont
  {Zhang}}, \bibinfo {author} {\bibfnamefont {A.}~\bibnamefont {Sit}}, \bibinfo
  {author} {\bibfnamefont {F.}~\bibnamefont {Bouchard}}, \bibinfo {author}
  {\bibfnamefont {H.}~\bibnamefont {Larocque}}, \bibinfo {author}
  {\bibfnamefont {F.}~\bibnamefont {Grenapin}}, \bibinfo {author}
  {\bibfnamefont {E.}~\bibnamefont {Cohen}}, \bibinfo {author} {\bibfnamefont
  {A.~C.}\ \bibnamefont {Elitzur}}, \bibinfo {author} {\bibfnamefont {J.~L.}\
  \bibnamefont {Harden}}, \bibinfo {author} {\bibfnamefont {R.~W.}\
  \bibnamefont {Boyd}}, \ and\ \bibinfo {author} {\bibfnamefont
  {E.}~\bibnamefont {Karimi}},\ }\href@noop {} {\bibfield  {journal} {\bibinfo
  {journal} {Optics express}\ }\textbf {\bibinfo {volume} {27}},\ \bibinfo
  {pages} {2212} (\bibinfo {year} {2019})}\BibitemShut {NoStop}%
\bibitem [{\citenamefont {Mitchison}\ and\ \citenamefont
  {Jozsa}(2001)}]{mitchison2001counterfactual}%
  \BibitemOpen
  \bibfield  {author} {\bibinfo {author} {\bibfnamefont {G.}~\bibnamefont
  {Mitchison}}\ and\ \bibinfo {author} {\bibfnamefont {R.}~\bibnamefont
  {Jozsa}},\ }\href@noop {} {\bibfield  {journal} {\bibinfo  {journal}
  {Proceedings of the Royal Society of London. Series A: Mathematical, Physical
  and Engineering Sciences}\ }\textbf {\bibinfo {volume} {457}},\ \bibinfo
  {pages} {1175} (\bibinfo {year} {2001})}\BibitemShut {NoStop}%
\bibitem [{\citenamefont {Hosten}\ \emph {et~al.}(2006)\citenamefont {Hosten},
  \citenamefont {Rakher}, \citenamefont {Barreiro}, \citenamefont {Peters},\
  and\ \citenamefont {Kwiat}}]{hosten2006counterfactual}%
  \BibitemOpen
  \bibfield  {author} {\bibinfo {author} {\bibfnamefont {O.}~\bibnamefont
  {Hosten}}, \bibinfo {author} {\bibfnamefont {M.~T.}\ \bibnamefont {Rakher}},
  \bibinfo {author} {\bibfnamefont {J.~T.}\ \bibnamefont {Barreiro}}, \bibinfo
  {author} {\bibfnamefont {N.~A.}\ \bibnamefont {Peters}}, \ and\ \bibinfo
  {author} {\bibfnamefont {P.~G.}\ \bibnamefont {Kwiat}},\ }\href@noop {}
  {\bibfield  {journal} {\bibinfo  {journal} {Nature}\ }\textbf {\bibinfo
  {volume} {439}},\ \bibinfo {pages} {949} (\bibinfo {year}
  {2006})}\BibitemShut {NoStop}%
\bibitem [{\citenamefont {Kong}\ \emph {et~al.}(2015)\citenamefont {Kong},
  \citenamefont {Ju}, \citenamefont {Huang}, \citenamefont {Wang},
  \citenamefont {Kong}, \citenamefont {Shi}, \citenamefont {Jiang},\ and\
  \citenamefont {Du}}]{kong2015experimental}%
  \BibitemOpen
  \bibfield  {author} {\bibinfo {author} {\bibfnamefont {F.}~\bibnamefont
  {Kong}}, \bibinfo {author} {\bibfnamefont {C.}~\bibnamefont {Ju}}, \bibinfo
  {author} {\bibfnamefont {P.}~\bibnamefont {Huang}}, \bibinfo {author}
  {\bibfnamefont {P.}~\bibnamefont {Wang}}, \bibinfo {author} {\bibfnamefont
  {X.}~\bibnamefont {Kong}}, \bibinfo {author} {\bibfnamefont {F.}~\bibnamefont
  {Shi}}, \bibinfo {author} {\bibfnamefont {L.}~\bibnamefont {Jiang}}, \ and\
  \bibinfo {author} {\bibfnamefont {J.}~\bibnamefont {Du}},\ }\href@noop {}
  {\bibfield  {journal} {\bibinfo  {journal} {Physical review letters}\
  }\textbf {\bibinfo {volume} {115}},\ \bibinfo {pages} {080501} (\bibinfo
  {year} {2015})}\BibitemShut {NoStop}%
\bibitem [{\citenamefont {Vaidman}(2003)}]{vaidman2003meaning}%
  \BibitemOpen
  \bibfield  {author} {\bibinfo {author} {\bibfnamefont {L.}~\bibnamefont
  {Vaidman}},\ }\href@noop {} {\bibfield  {journal} {\bibinfo  {journal}
  {Foundations of Physics}\ }\textbf {\bibinfo {volume} {33}},\ \bibinfo
  {pages} {491} (\bibinfo {year} {2003})}\BibitemShut {NoStop}%
\bibitem [{\citenamefont {Simon}\ and\ \citenamefont
  {Platzman}(2000)}]{simon2000fundamental}%
  \BibitemOpen
  \bibfield  {author} {\bibinfo {author} {\bibfnamefont {S.~H.}\ \bibnamefont
  {Simon}}\ and\ \bibinfo {author} {\bibfnamefont {P.}~\bibnamefont
  {Platzman}},\ }\href@noop {} {\bibfield  {journal} {\bibinfo  {journal}
  {Physical Review A}\ }\textbf {\bibinfo {volume} {61}},\ \bibinfo {pages}
  {052103} (\bibinfo {year} {2000})}\BibitemShut {NoStop}%
\bibitem [{\citenamefont {Dicke}(1986)}]{dicke1986observing}%
  \BibitemOpen
  \bibfield  {author} {\bibinfo {author} {\bibfnamefont {R.}~\bibnamefont
  {Dicke}},\ }\href@noop {} {\bibfield  {journal} {\bibinfo  {journal}
  {Foundations of physics}\ }\textbf {\bibinfo {volume} {16}},\ \bibinfo
  {pages} {107} (\bibinfo {year} {1986})}\BibitemShut {NoStop}%
\bibitem [{\citenamefont {Feynman}\ \emph {et~al.}(2010)\citenamefont
  {Feynman}, \citenamefont {Hibbs},\ and\ \citenamefont
  {Styer}}]{feynman2010quantum}%
  \BibitemOpen
  \bibfield  {author} {\bibinfo {author} {\bibfnamefont {R.~P.}\ \bibnamefont
  {Feynman}}, \bibinfo {author} {\bibfnamefont {A.~R.}\ \bibnamefont {Hibbs}},
  \ and\ \bibinfo {author} {\bibfnamefont {D.~F.}\ \bibnamefont {Styer}},\
  }\href@noop {} {\bibfield  {journal} {\bibinfo  {journal} {Mineola, NY:
  Dover}\ } (\bibinfo {year} {2010})}\BibitemShut {NoStop}%
\bibitem [{\citenamefont {Feynman}(2006)}]{feynman2006qed}%
  \BibitemOpen
  \bibfield  {author} {\bibinfo {author} {\bibfnamefont {R.~P.}\ \bibnamefont
  {Feynman}},\ }\href@noop {} {\emph {\bibinfo {title} {QED: The strange theory
  of light and matter}}}\ (\bibinfo  {publisher} {Princeton University Press},\
  \bibinfo {year} {2006})\BibitemShut {NoStop}%
\bibitem [{\citenamefont {Dressel}\ \emph {et~al.}(2013)\citenamefont
  {Dressel}, \citenamefont {Lyons}, \citenamefont {Jordan}, \citenamefont
  {Graham},\ and\ \citenamefont {Kwiat}}]{dressel2013strengthening}%
  \BibitemOpen
  \bibfield  {author} {\bibinfo {author} {\bibfnamefont {J.}~\bibnamefont
  {Dressel}}, \bibinfo {author} {\bibfnamefont {K.}~\bibnamefont {Lyons}},
  \bibinfo {author} {\bibfnamefont {A.~N.}\ \bibnamefont {Jordan}}, \bibinfo
  {author} {\bibfnamefont {T.~M.}\ \bibnamefont {Graham}}, \ and\ \bibinfo
  {author} {\bibfnamefont {P.~G.}\ \bibnamefont {Kwiat}},\ }\href@noop {}
  {\bibfield  {journal} {\bibinfo  {journal} {Physical Review A}\ }\textbf
  {\bibinfo {volume} {88}},\ \bibinfo {pages} {023821} (\bibinfo {year}
  {2013})}\BibitemShut {NoStop}%
\bibitem [{\citenamefont {Goodman}(2005)}]{goodman2005introduction}%
  \BibitemOpen
  \bibfield  {author} {\bibinfo {author} {\bibfnamefont {J.~W.}\ \bibnamefont
  {Goodman}},\ }\href@noop {} {\emph {\bibinfo {title} {Introduction to Fourier
  optics}}}\ (\bibinfo  {publisher} {Roberts and Company Publishers},\ \bibinfo
  {year} {2005})\BibitemShut {NoStop}%
\bibitem [{\citenamefont {Sawant}\ \emph {et~al.}(2014)\citenamefont {Sawant},
  \citenamefont {Samuel}, \citenamefont {Sinha}, \citenamefont {Sinha},\ and\
  \citenamefont {Sinha}}]{sawant2014nonclassical}%
  \BibitemOpen
  \bibfield  {author} {\bibinfo {author} {\bibfnamefont {R.}~\bibnamefont
  {Sawant}}, \bibinfo {author} {\bibfnamefont {J.}~\bibnamefont {Samuel}},
  \bibinfo {author} {\bibfnamefont {A.}~\bibnamefont {Sinha}}, \bibinfo
  {author} {\bibfnamefont {S.}~\bibnamefont {Sinha}}, \ and\ \bibinfo {author}
  {\bibfnamefont {U.}~\bibnamefont {Sinha}},\ }\href@noop {} {\bibfield
  {journal} {\bibinfo  {journal} {Physical review letters}\ }\textbf {\bibinfo
  {volume} {113}},\ \bibinfo {pages} {120406} (\bibinfo {year}
  {2014})}\BibitemShut {NoStop}%
\bibitem [{\citenamefont {Beau}(2012)}]{beau2012feynman}%
  \BibitemOpen
  \bibfield  {author} {\bibinfo {author} {\bibfnamefont {M.}~\bibnamefont
  {Beau}},\ }\href@noop {} {\bibfield  {journal} {\bibinfo  {journal} {European
  Journal of Physics}\ }\textbf {\bibinfo {volume} {33}},\ \bibinfo {pages}
  {1023} (\bibinfo {year} {2012})}\BibitemShut {NoStop}%
\bibitem [{\citenamefont {Broadbent}\ \emph {et~al.}(2009)\citenamefont
  {Broadbent}, \citenamefont {Zerom}, \citenamefont {Shin}, \citenamefont
  {Howell},\ and\ \citenamefont {Boyd}}]{broadbent2009discriminating}%
  \BibitemOpen
  \bibfield  {author} {\bibinfo {author} {\bibfnamefont {C.~J.}\ \bibnamefont
  {Broadbent}}, \bibinfo {author} {\bibfnamefont {P.}~\bibnamefont {Zerom}},
  \bibinfo {author} {\bibfnamefont {H.}~\bibnamefont {Shin}}, \bibinfo {author}
  {\bibfnamefont {J.~C.}\ \bibnamefont {Howell}}, \ and\ \bibinfo {author}
  {\bibfnamefont {R.~W.}\ \bibnamefont {Boyd}},\ }\href@noop {} {\bibfield
  {journal} {\bibinfo  {journal} {Physical Review A}\ }\textbf {\bibinfo
  {volume} {79}},\ \bibinfo {pages} {033802} (\bibinfo {year}
  {2009})}\BibitemShut {NoStop}%
\bibitem [{\citenamefont {Misra}\ and\ \citenamefont
  {Sudarshan}(1977)}]{misra1977zeno}%
  \BibitemOpen
  \bibfield  {author} {\bibinfo {author} {\bibfnamefont {B.}~\bibnamefont
  {Misra}}\ and\ \bibinfo {author} {\bibfnamefont {E.~G.}\ \bibnamefont
  {Sudarshan}},\ }\href@noop {} {\bibfield  {journal} {\bibinfo  {journal}
  {Journal of Mathematical Physics}\ }\textbf {\bibinfo {volume} {18}},\
  \bibinfo {pages} {756} (\bibinfo {year} {1977})}\BibitemShut {NoStop}%
\bibitem [{\citenamefont {Mackrory}\ \emph {et~al.}(2010)\citenamefont
  {Mackrory}, \citenamefont {Jacobs},\ and\ \citenamefont
  {Steck}}]{mackrory2010reflection}%
  \BibitemOpen
  \bibfield  {author} {\bibinfo {author} {\bibfnamefont {J.~B.}\ \bibnamefont
  {Mackrory}}, \bibinfo {author} {\bibfnamefont {K.}~\bibnamefont {Jacobs}}, \
  and\ \bibinfo {author} {\bibfnamefont {D.~A.}\ \bibnamefont {Steck}},\
  }\href@noop {} {\bibfield  {journal} {\bibinfo  {journal} {New Journal of
  Physics}\ }\textbf {\bibinfo {volume} {12}},\ \bibinfo {pages} {113023}
  (\bibinfo {year} {2010})}\BibitemShut {NoStop}%
\bibitem [{\citenamefont {Halliwell}\ and\ \citenamefont
  {Yearsley}(2012)}]{halliwell2012pitfalls}%
  \BibitemOpen
  \bibfield  {author} {\bibinfo {author} {\bibfnamefont {J.}~\bibnamefont
  {Halliwell}}\ and\ \bibinfo {author} {\bibfnamefont {J.}~\bibnamefont
  {Yearsley}},\ }\href@noop {} {\bibfield  {journal} {\bibinfo  {journal}
  {Physical Review D}\ }\textbf {\bibinfo {volume} {86}},\ \bibinfo {pages}
  {024016} (\bibinfo {year} {2012})}\BibitemShut {NoStop}%
\end{thebibliography}%
\end{document}